\documentclass[aps,prl,reprint,amsmath,amssymb,amsfonts,floatfix, showpacs,intlimits, euspript,superscriptaddress]{revtex4-2}

\usepackage{graphicx}
\usepackage{dcolumn}
\usepackage{bm}
\usepackage{hyperref}

\usepackage{xcolor,soul}
\hypersetup{
    colorlinks=true,
    linkcolor=blue,
    citecolor=blue,
    urlcolor=cyan,
}

\usepackage{xcolor}

\begin{document}

\preprint{APS/123-QED}

\title{Femtosecond laser comb driven perpendicular standing spin waves} 

\author{A. A. Awad}
\email{ahmad.awad@physics.gu.se}
\affiliation{Physics Department, University of Gothenburg, 412 96 Gothenburg, Sweden.}

\author{S. Muralidhar}
\affiliation{Physics Department, University of Gothenburg, 412 96 Gothenburg, Sweden.}
\author{A. Alem\'an}
\affiliation{Physics Department, University of Gothenburg, 412 96 Gothenburg, Sweden.}

\author{R. Khymyn}
\affiliation{Physics Department, University of Gothenburg, 412 96 Gothenburg, Sweden.}

\author{D. Hanstorp}
\affiliation{Physics Department, University of Gothenburg, 412 96 Gothenburg, Sweden.}
\author{J. \AA kerman}
\affiliation{Physics Department, University of Gothenburg, 412 96 Gothenburg, Sweden.}

\date{\today}

\begin{abstract}

We study femtosecond laser comb driven sustained and coherent spin wave (SW) generation in permalloy (Py) films over a thickness range of $d =$ 40--100 nm. 
A simple rapid demagnetization model describes the dependence of the observed SW intensity on laser power for all film thicknesses. 
In the thicker films we observe laser comb excited perpendicular standing spin waves up to third order and to 18 multiples of the 1 GHz laser repetition rate. 
Our results demonstrate the versatility of femtosecond combs as contact-less SW point sources over a wide range of film thickness and type of SW modes.

\end{abstract}

\maketitle
The recent rapid development of magnonics offers tremendous potential for 
promising applications using compact spin wave (SW) devices \cite{Kruglyak2010Magnonics,Demokritov2013Magnonicbook,Chumak2015Magnspin,Pirro2021NatRevMat} with fast, integrated, energy-efficient, and  CMOS compatible functionalities. Meanwhile, magnonics research has largely relied on radio-frequency antenna based excitation \cite{Bailleul2001EPL,Johnson2005APL,Demidov2009apl,Demidov2011apl,Pirro2014apl}, or more recently confined spin currents to excite \cite{Tsoi1998prl,Rippard2004apl,Demidov2010natmat,madami2011natnano, Dumas2014ieeetmag,Chen2016procieee}, control \cite{Demidov2011prl, Haidar2016prb, Balinsky2016ieeeml}, and readout \cite{Balinsky2015ieeeml, bracher2017nl} SWs. While it typically limits both the accessible frequency range and the energy efficiency, it also often requires substantial nano-fabrication to reach short SW wavelengths \cite{Barman2021Roadmap, Yu2016natcomm, houshang2018spin}.

Optical excitation of the spin degree of freedom, on the other hand, is so far the fastest and the most energy-efficient, as shown by femtosecond (fs) laser pulse magnetization switching, which just requires a few hundreds of aJ at less than 20 ps timescale \cite{Stupakiewicz2017Nature}. As nano-fabrication is not necessary, light-mediated excitation of SWs using femtosecond lasers  \cite{VanKampen2002,Kimel2005,Satoh2010_PRL,Satoh2012_NatPho,Au2013,Yun2015,Iihama2016b,Bossini2016,Hashimoto_2017_NatComm,Kamimaki2017,Hashimoto_2018_PRB,Rasing2010Rev,Bossini_2017,Cao2021nanoscale}
is hence very appealing for contact-less, ultra-fast and energy efficient SW generation. Even so, since SWs have a decay time on the order of nanoseconds, it is not possible to sustain continuous SW excitation. Furthermore, as fs-pulsed lasers typically have low repetition rates, it has not been meaningful to study the excited SWs with time-averaged techniques such as Brillouin light spectroscopy. 
Recently, frequency comb fs-lasers were employed for SW excitation \cite{Jackl2017,Savochkin2017,Kobecki2020NatComm}, resulting in much higher repetition rates. As a consequence, the SW dynamics can be sustained as long as it survives to the next pulse arrival, allowing for steady state continuous SW excitation by optical means. This powerful contact-less and patterning-free generation of SWs
has since been used to \emph{e.g.}~enhance Brillouin light scattering microscopy \cite{Muralidhar2019PRB,Aleman2020OSA} and create caustic SW beams through optical excitation \cite{Muralidhar2021PRL}. 
\begin{figure}[b]
    \begin{center}
        \includegraphics[width=0.92\linewidth]{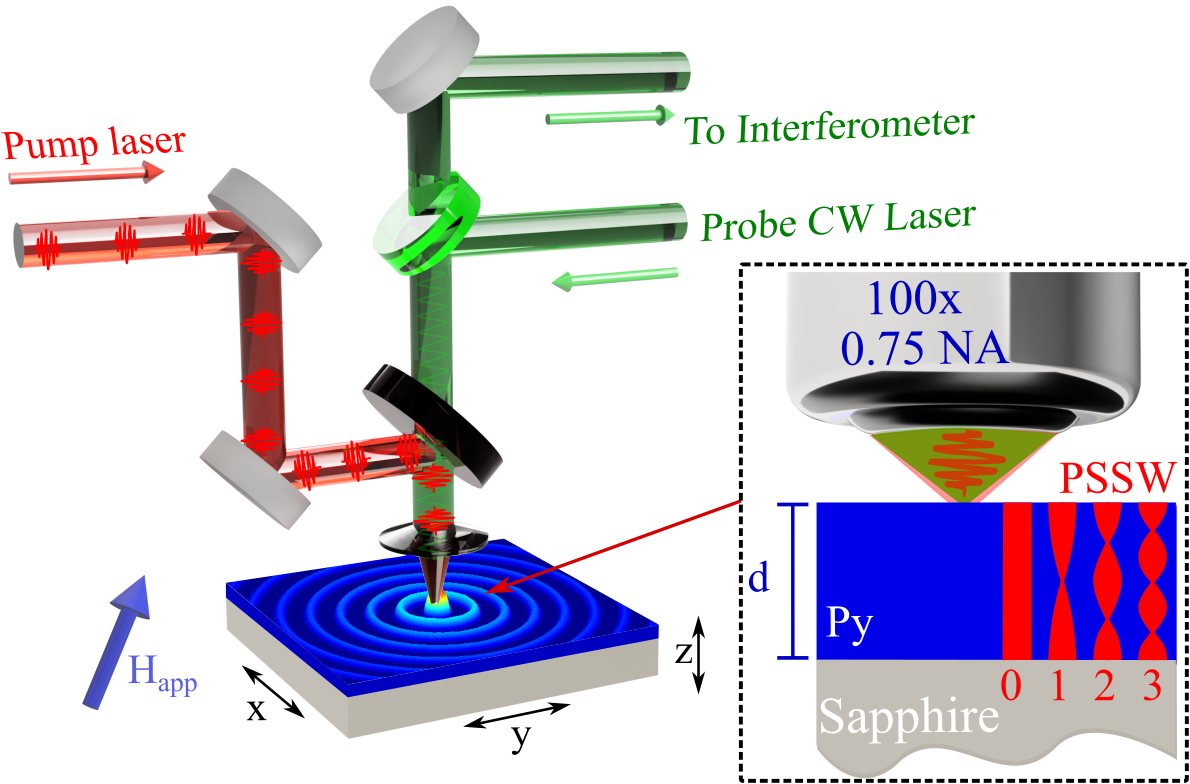}
        \caption{Schematic of the experiment.
        Py (NiFe) films with thickness $d =$ 40--100 nm are pumped with a 1 GHz-pulsed femtosecond laser (120 fs) at a wavelength of  816 nm (red) and probed by a continuous wave single-mode 532 nm BLS laser (green). The magnetic field $\mathit{H_{app}}$ is applied at an 82$^\circ$ out-of-plane angle with its in-plane component along the $y$ axis. The dashed box shows the sample below the objective with instantaneous profiles of the first three perpendicular standing spinwaves (PSSW; red regions).}
        \label{fig:fig1}
    \end{center}
\end{figure}

In this work we study fs-laser frequency comb generated SWs in thicker, $d =$ 40--100 nm, Permalloy (Py: Ni$_{80}$Fe$_{20}$) films and explore whether we can excite additional SW modes such as quantized perpendicular standing SWs (PSSWs). In the thicker films, we indeed find that up to the first three PSSW modes can be excited by the fs-laser comb, and that the first mode shows a higher efficiency of excitation than that of the ferromagnetic resonance (FMR). The higher PSSW modes show a much weaker intensity and the fs-laser drive excitation is barely visible for PSSW3.
Permalloy (Py) thin films were fabricated using magnetron sputtering onto 0.5 mm thick $c$-plane  sapphire substrates (2 $\times$ 2 cm$^2$). Sapphire was chosen for its excellent heat conductivity to ensure efficient dissipation of the laser generated heat. 
The Py film thickness was varied from 40 to 100 nm, with a 4 nm SiO$x$ terminating layer to prevent oxidation. Figure~\ref{fig:fig1} shows a schematic of the experiment. The frequency comb enhanced BLS microscope combines a 120~fs, 1 GHz repetition rate, mode-locked laser pump operating at 816~nm with a BLS single mode laser probe at 532 nm. Both lasers are focused onto the Py films down to their respective diffraction limits using a high numerical aperture (NA = 0.75) objective lens. A more in-depth description of the set-up can be found in Ref. \cite{Aleman2020OSA}.

\begin{figure}[b]
    \centering
    \includegraphics[width=\linewidth]{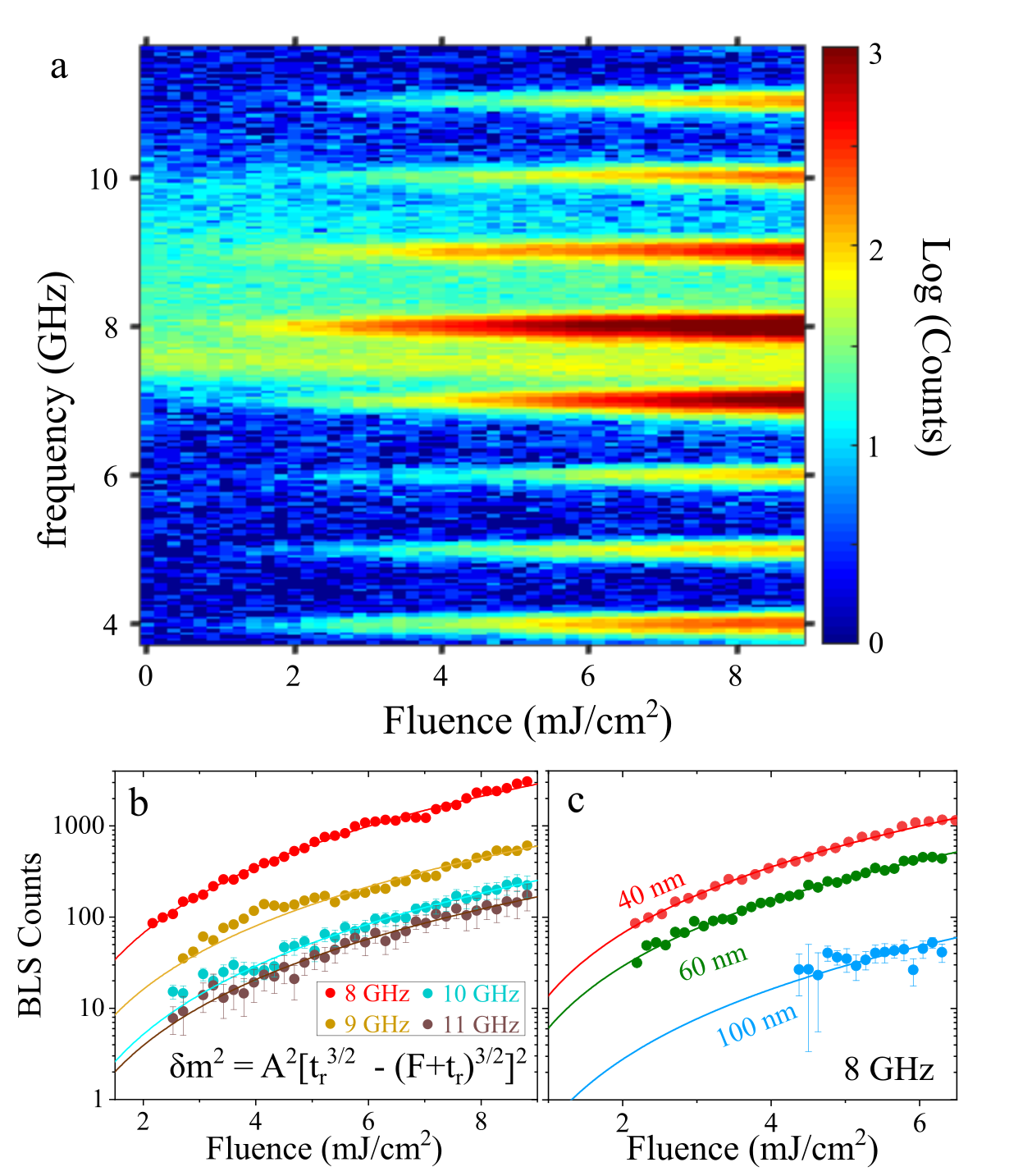}
    \caption{(a) SW intensity (log BLS counts) of a 40 nm Py film as a function of laser fluence. (b) Peak BLS counts extracted from (a) at 8, 9, 10 and 11 GHz with solid lines being fits to Eq.~\ref{eq:Bloch_fit}. (c) Peak BLS counts \emph{vs.}~laser fluence for the FMR modes in three different Py films with $d$ = 40, 60, and 100 nm, with fits to
    Eq.~\eqref{eq:Bloch_fit}}
    \label{fig:fig2}
\end{figure}

Figure~\ref{fig:fig2}(a) shows a log plot of frequency-resolved BLS counts in the 40 nm Py film in an obliquely applied field of 600 mT \emph{vs.}~the fluence of the fs-laser excitation. Without any laser excitation, the BLS spectrum shows the usual SW gap at about 7.5 GHz, where the BLS counts rapidly increase due to  FMR, and then a gradually decreasing BLS count at higher frequency corresponding to the thermal distribution of SWs. At about a fluence of 2 mJ/cm$^2$, additional SW peaks begin to appear at the higher harmonics of the 1 GHz fs-laser repetition rate. While the response is dominated by peaks closest to the FMR, the log scale resolves additional counts at all harmonics, clearly showing that both localized and propagating SWs can be excited to different degrees. The excited SWs show a similar trend as in previous reports \cite{Muralidhar2019PRB} where the SWs having a frequency of a multiple of the frequency comb repetition rate are coherently amplified as a consequence of rapid demagnetization, with each new arriving pulse adding magnons in-phase with the existing generated magnon population not yet dissipated from the previous pump pulse. 
 To model the laser fluence dependence of the observed BLS counts, we assume that each laser pulse causes short term heating of the magnetic subsystem, which in turn changes the demagnetizing field $\mathit{H_{dem}}=-M_s$. For the temperatures $T$ far enough from Curie point $T_c$ the magnetization should follow the Bloch law with the change of magnetization as $\Delta M_s=\frac{M_0}{T_c^{3/2}}[T^{3/2}-(T+\Delta T)]^{3/2}$, where $M_0$ is the magnetization at zero temperature, and $T$ is close to room temperature, because of the high efficiency of the heat dissipation through the substrate. Assuming that the heating is proportional to the laser fluence $\Delta T=\alpha F$ (where $\alpha$ defines the efficiency of heating), one can rewrite the above as $\Delta M=\frac{M_0 \alpha^{3/2}}{T_c^{3/2}}\left[\left(\frac{T}{\alpha}\right)^{\frac{3}{2}}-\left(\frac{T}{\alpha}+F\right)^{\frac{3}{2}}\right]$. Taking into account that the power, absorbed by a resonance mode, or in other words the magnon population is quadratic in the excitation field, and denoting $t_r=T/\alpha$,  the BLS counts vs.~fluence can be well fitted [Fig.~\ref{fig:fig2}(b)] by \cite{Muralidhar2019PRB}
\begin{equation}
    \delta M_{\text{s}}^2 = A^2(t_\text{r}^{3/2} -  (t_\text{r} +F)^{3/2})^2, 
    \label{eq:Bloch_fit}
\end{equation}
where $A \propto M_0 (\alpha/T_c)^{\frac{3}{2}}$ with the coefficient defined by the anti-Hermitian part of magnetic susceptibility tensor \cite{akhiezer1968spin}.

Measurements at the same conditions of field, frequency range, and fs-laser fluence range were then conducted on the thicker Py films. 
As varying the film thickness will slightly change the in-plane wavevector frequencies, particular consideration should be taken into account to correctly compare SWs generated at different thicknesses. Since the FMR peaks are located at similar frequencies for all films, we chose the 8 GHz BLS peak just above FMR as a strong and comparable mode between different thicknesses. Fig.~\ref{fig:fig2}(c) compares the fs-laser generated response for all film thicknesses. Again, the response vs.~fs-laser fluence can be very well fit to Eq.~\ref{eq:Bloch_fit} for all thicknesses, 
clearly indicating the efficiency and the versatility of using a high frequency fs-laser comb to sustain SWs over a wide thickness range. 
\begin{figure}[b]
    \centering
    \includegraphics[width=\linewidth]{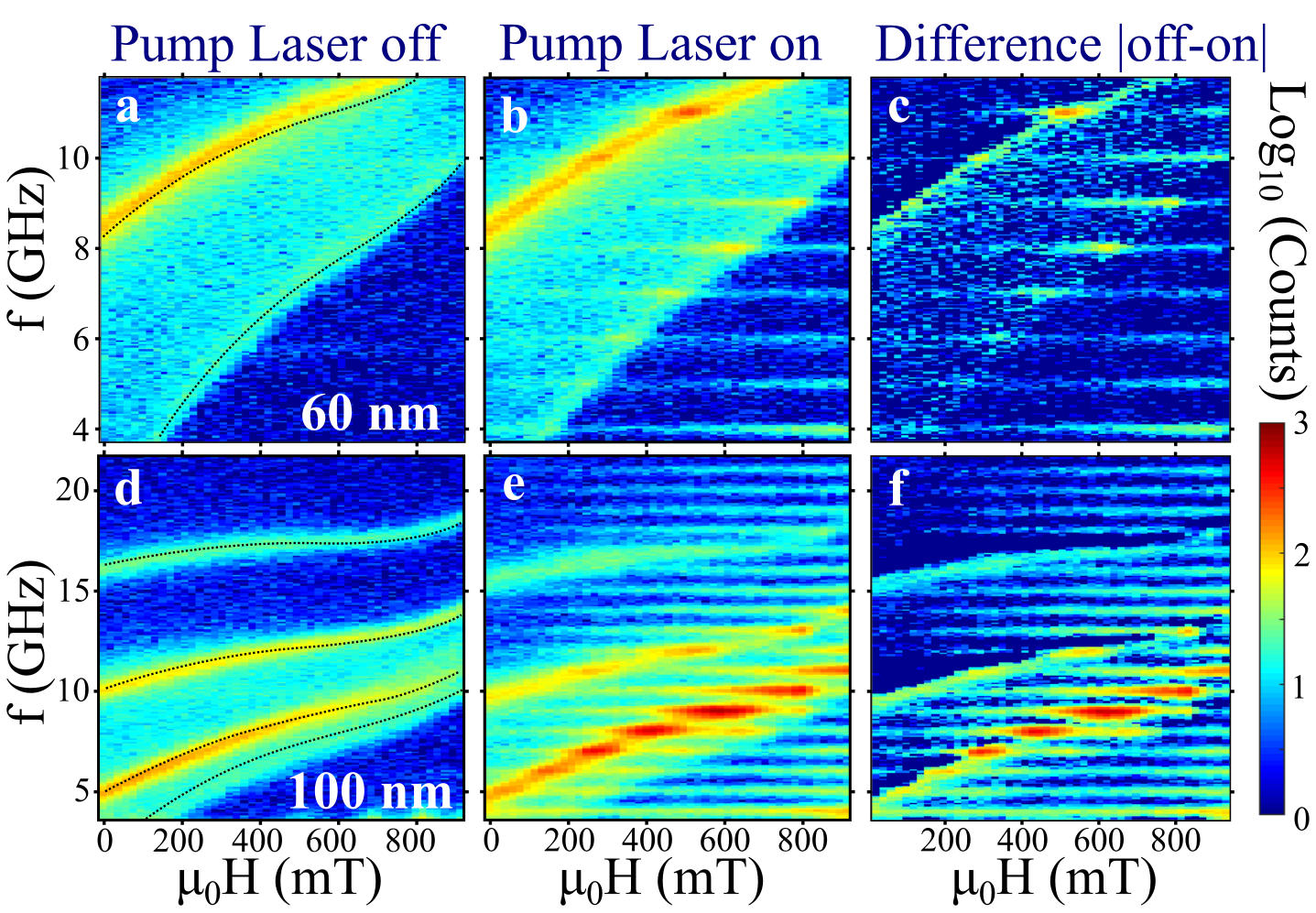}
    \caption{ 
    (a),(d) $\mu$-BLS counts (log scale) \emph{vs.}~applied magnetic field for 60 nm and 100 nm, respectively; black dotted lines are fits to the dispersion relation, Eq.~\ref{eq:PSSWdisp}. (b),(e) SW spectrum at a fs-laser fluence of 2.5 and 5 $\mathrm{mJ/cm^2}$ for 60 nm and 100 nm films, respectively.
    (c) The difference between spectra in (a) and (b). (f) The difference between spectra in (d) and (e). 
    }
    \label{fig:fig4}
\end{figure}

A particular benefit of frequency comb enhanced BLS microscopy is the very high frequency content of the comb (upper limit given by the 120 fs pulse length) allowing the simultaneous exploration of a large number of SW modes to very high frequencies.
In principle, any harmonic of the comb can drive a corresponding SW mode into a coherent steady state, provided the excitation efficiency is sufficient and the SW damping low enough to allow enough magnons to survive to the next pulse before being damped out. 
The PSSW modes of the thicker films here offer us the opportunity to demonstrate that they, too, can be driven coherently by the frequency comb. 
Fig.~\ref{fig:fig4} shows the thermal SW spectra from the (a) 60 and (d) 100 nm Py films \emph{vs.}~field magnitude before the application of the fs-laser pump. The SW gap and the FMR peak are clearly seen at the low frequency range and then pronounced additional peaks, related to the PSSW modes, can be observed at higher frequency.  
From these measurements, the SW frequencies are extracted by fitting the BLS spectra to multiple Lorentzians. The extracted frequencies are then fitted to the calculated magnon dispersion using a modified version of Eq.~(45) in  \cite{kalinikos1986theory}. For the case of PSSWs, which have zero in-plane vector and are quantized along the film thickness, the dispersion relation reads: 
\begin{equation}
\begin{split}
    f = \frac{\gamma \mu_0}{2\pi} \sqrt{H_{int}+M_s l_{ex}^2 q^2} \\ \sqrt{H_{int}+M_s l_{ex}^2 q^2+M_s \cos^2\theta_{in}}
        \label{eq:PSSWdisp}
\end{split}
\end{equation}
where $l_{ex}=\sqrt{2 A_{ex}/(\mu_0 M_s^2)}$ is the exchange length, $q= p\pi/d$ is the out-of-plane (OOP) wavevector, $p$ represents the integer of a half wavelength along the film thickness, and $\mathit{H_{int}}$ and $\theta_{int}$ are the magnitude and OOP angle of the internal field, calculated from solving the magnetostatic problem (see \emph{e.g.} Eq.~2.3 in  \cite{houshang2018spin}):
\begin{eqnarray}
H_{app}\cos \theta_{app} &=& H_{\mathrm{int}} \cos \theta_{\mathrm{int}} \\
H_{app}\sin \theta_{app} &=& (H_{\mathrm{int}} + M_{\mathrm{s}}) \sin \theta_{\mathrm{int}},
\end{eqnarray} 
where $H_{app}$ denotes the external magnetic field applied at an OOP angle $\theta_{app}$. Fits to Eq.~2--4 are shown as lines in Fig.~\ref{fig:fig4}(a) and (d), with the extracted material parameters:  saturation magnetization $\mu_0 M_s=0.82 \pm 0.01~\text{T}$, exchange stiffness $A_{ex}=10.1 \pm 0.04~\text{pJ/m}$, gyromagnetic ratio
 $\gamma/2\pi=29.4 \pm 0.1$, all largely reproducible for all film thicknesses.

The laser-induced SWs in Fig.~\ref{fig:fig4}(b) and (e) exhibit a similar field dependence as the thermally generated SWs in Fig.~\ref{fig:fig4}(a) and we find that all observed PSSW modes can be driven by the frequency comb. To better bring out this result we subtract the thermal spectra in Fig.~\ref{fig:fig4}(a) and (d) from the thermal+laser spectra in Fig.~\ref{fig:fig4}(b) and (e) and plot the difference in Fig.~\ref{fig:fig4}(c) and (f), which clearly show the presence of fs-laser induced SWs across all modes. We first observe that the frequency comb driven contribution is much greater for PSSW1 than for the FMR mode, in particular in the 100 nm film. This may result from the thermal PSSW1 mode showing more intensity to begin with but could also reflect a preference to energize the non-uniform PSSW1 mode if the laser pulse is primarily absorbed in the upper part of the thicker films. In contrast, the two higher PSSW modes are much less excited and PSSW3 is just barely visible, which is consistent with the higher SW damping at higher frequency. Nevertheless, we can observed additional BLS counts in the PSSW3 mode up to the $18^{th}$ harmonic of the frequency comb. 

In conclusion, we have used $\mu$-BLS microscopy to study fs-laser frequency comb driven coherent SW excitations in Py films over a large thickness range. We find that the SW intensity is consistent with a simple model of the absorbed energy density determining the local demagnetization as given by the temperature of the magnon bath. 
Taking advantage of the detection of zero in-plane vector perpendicular standing spin waves (PSSW) in thick films, which are not limited by the wavevector resolution in $\mu$-BLS, we clearly observed sustained coherent SWs on the first 18 multiples of the 1 GHz repetition rate of the frequency comb. The efficient excitation at such high frequencies compared to the fundamental repetition rate highlights the potential and versatility of this approach. As lasers with yet higher repetition rates are now available, future studies should likely be able to coherently excite much higher frequency SW modes in ferro-, ferri-, and potentially antiferromagnetic thin films, for both fundamental research and applications.

\emph{Acknowledgments} Students M. Toft{\aa}s,  O. Lundgren, N. Tornberg, I. Rydbjer, H. Bergstr{\"o}m,  M. Samuelsson are gratefully acknowledged for their participation in the early stages of the experiments as part of their Bachelor theses. This work was partially supported by the Swedish Research Council (VR), the Knut and Alice Wallenberg foundation (KAW), and the Horizon 2020 research and innovation programme (ERC Advanced Grant No.~835068 "TOPSPIN").
\section*{Data availability}
The data that support the findings of this study are available from the corresponding author upon reasonable request.

\bibliography{references}

\end{document}